\documentclass[prl,twocolumn,amsmath,amssymb,superscriptaddress, showpacs]{revtex4-1}

\usepackage{amsmath}    % need for subequations
\usepackage{graphicx}   % need for figures %dblfloatfix
\usepackage{verbatim}   % useful for program listings
\usepackage{color}      % use if color is used in text
\usepackage[usenames,dvipsnames,svgnames,table]{xcolor}
\usepackage{subfigure}  % use for side-by-side figures
\usepackage[hidelinks]{hyperref}   % use for hypertext links, including those to external documents and URLs
\usepackage{multirow}
\usepackage{float}
\usepackage{braket}
\usepackage{siunitx}
\usepackage{bbm}
\usepackage{microtype}       % als letztes -- Verbesserter Textsatz
\usepackage{bibunits}

\usepackage{ulem}
\renewcommand{\emph}[1]{\textit{#1}}

\defaultbibliography{fftcd_library_v7}

\graphicspath{{images/}}
\newcommand{\ind}[1]{\ensuremath{_{\text{#1}}}}
\newcommand{\dni}[1]{\ensuremath{^{\text{#1}}}}
\newcommand{\inds}[1]{\ensuremath{_{\textsf{#1}}}}
\newcommand{\dnis}[1]{\ensuremath{^{\textsf{#1}}}}
\let\shorti\i
\renewcommand{\i}{\mathrm i}
\newcommand{\unit}[1]{\ensuremath{\,\mathrm{#1}}}
\newcommand{\e}[1]{\ensuremath{\mathrm e^{#1}}}

\renewcommand{\dag}{^{\dagger}}
\newcommand{\dags}{^{\dagger 2}}
\newcommand{\dagc}{^{\dagger 3}}
\newcommand{\dagt}{^{\dagger 4}}
\newcommand{\colordc}{black}%{RoyalBlue}
\newcommand{\colord}{black}%{cyan}
\newcommand{\colordd}{black}%{red}
\newcommand{\colortwo}{black}
\newcommand{\colorthree}{black}
\newcommand{\colorsum}{black}%{ForestGreen}
\newcommand{\puttitle}{Observation of the Crossover from Photon Ordering to Delocalization\\in Tunably Coupled Resonators}

\begin{document}

\title{\puttitle}
\author{Michele C. Collodo}
\email{michele.collodo@phys.ethz.ch}
\author{Anton Poto\v{c}nik}
\author{Simone Gasparinetti}
\author{Jean-Claude Besse}
\author{Marek~Pechal}
\altaffiliation{Current address: Ginzton Laboratory, Stanford University, {Stanford, California 94305}, USA}
\affiliation{Department of Physics, ETH Zurich, CH-8093 Zurich, Switzerland}

\author{Mahdi Sameti}
\author{Michael J. Hartmann}
\affiliation{Institute of Photonics and Quantum Sciences, Heriot-Watt University Edinburgh EH14 4AS, United Kingdom}

\author{Andreas Wallraff}
\author{Christopher Eichler}
\email{eichlerc@phys.ethz.ch}
\affiliation{Department of Physics, ETH Zurich, CH-8093 Zurich, Switzerland}

\date{\today}

%\pacs{\textsl{version 3.5, supplementary v2.2}}

\begin{abstract}
Networks of nonlinear resonators offer intriguing perspectives as quantum simulators for non-equilibrium many-body phases of driven-dissipative systems.
Here, we employ photon correlation measurements to study the radiation fields emitted from a system of two superconducting resonators, coupled nonlinearly by a superconducting quantum interference device (SQUID).
We apply a parametrically modulated magnetic flux to control the linear photon hopping rate between the two resonators and its ratio with the cross-Kerr rate.
When increasing the hopping rate,
we observe a crossover from an ordered to a delocalized state of photons.
The presented coupling scheme is intrinsically robust to frequency disorder and may therefore prove useful for realizing larger-scale resonator arrays.
\end{abstract}

\maketitle

%%%%%%%%%%%%%%%%%%%%%%%%%%%%%%%%%
\begin{bibunit}[apsrev4-1]

Engineering optical nonlinearities that are appreciable on the single photon level and lead to nonclassical light fields has been a central objective for the study of light-matter interaction in quantum optics \cite{haroche2013exploring, raimond_manipulating_2001, chang_quantum_2014}.
While such nonlinearities have first been realized in individual optical cavities \cite{thompson_observation_1992, birnbaum_photon_2005}
and with Rydberg atoms \cite{brune_quantum_1996, peyronel_quantum_2012},
more recently superconducting circuit quantum electrodynamics (QED) \cite{wallraff_strong_2004}
has proven to be a powerful platform for the study of nonclassical light fields. Circuit QED systems facilitate strong effective interactions between individual photons \cite{houck_generating_2007, lang_observation_2011},
long coherence times \cite{koch_charge-insensitive_2007}
as well as precise control of drive fields \cite{motzoi_simple_2009, heeres_implementing_2017}
within a large variety of possible design implementations.
Particularly, \textsl{in-situ} tunable or nonlinear couplers have been explored more recently for superconducting elements \cite{bertet_parametric_2006, baust_tunable_2015, mckay_universal_2016, chen_qubit_2014, lu_universal_2017, kounalakis_tuneable_2018, eichler_realizing_2018}.

Well-controllable engineered quantum systems, in which strong optical nonlinearities occur in extended volumes \cite{carusotto_quantum_2013} or networks of multiple nonlinear resonators offer interesting perspectives to study interacting many-body systems with photons \cite{hartmann_quantum_2016, noh_quantum_2017}
and to mimic the dynamics of otherwise less accessible systems \cite{georgescu_quantum_2014, schmidt_circuit_2013, houck_-chip_2012}, such as supersolids \cite{leonard_supersolid_2017} or topological quantum matter \cite{gross_quantum_2017}.
Photons are trapped in resonators only for a limited time, even in high quality devices. Interacting photons are thus typically explored in a non-equilibrium regime, in which continuous driving compensates for excitation loss and yields stationary states of light fields \cite{noh_out--equilibrium_2017}.
It has been predicted that these non-equilibrium systems offer rich phase diagrams of novel exotic states which have no analogue in equilibrium systems \cite{prosen_quantum_2008}, featuring \textsl{e.g.} synchronization \cite{leib_synchronized_2014} or bistability \cite{le_boite_steady-state_2013,  kessler_dissipative_2012}.

%%%%%%%%%%%
\begin{figure}[b!] % !b H
\centering
\includegraphics[width = 0.48\textwidth]{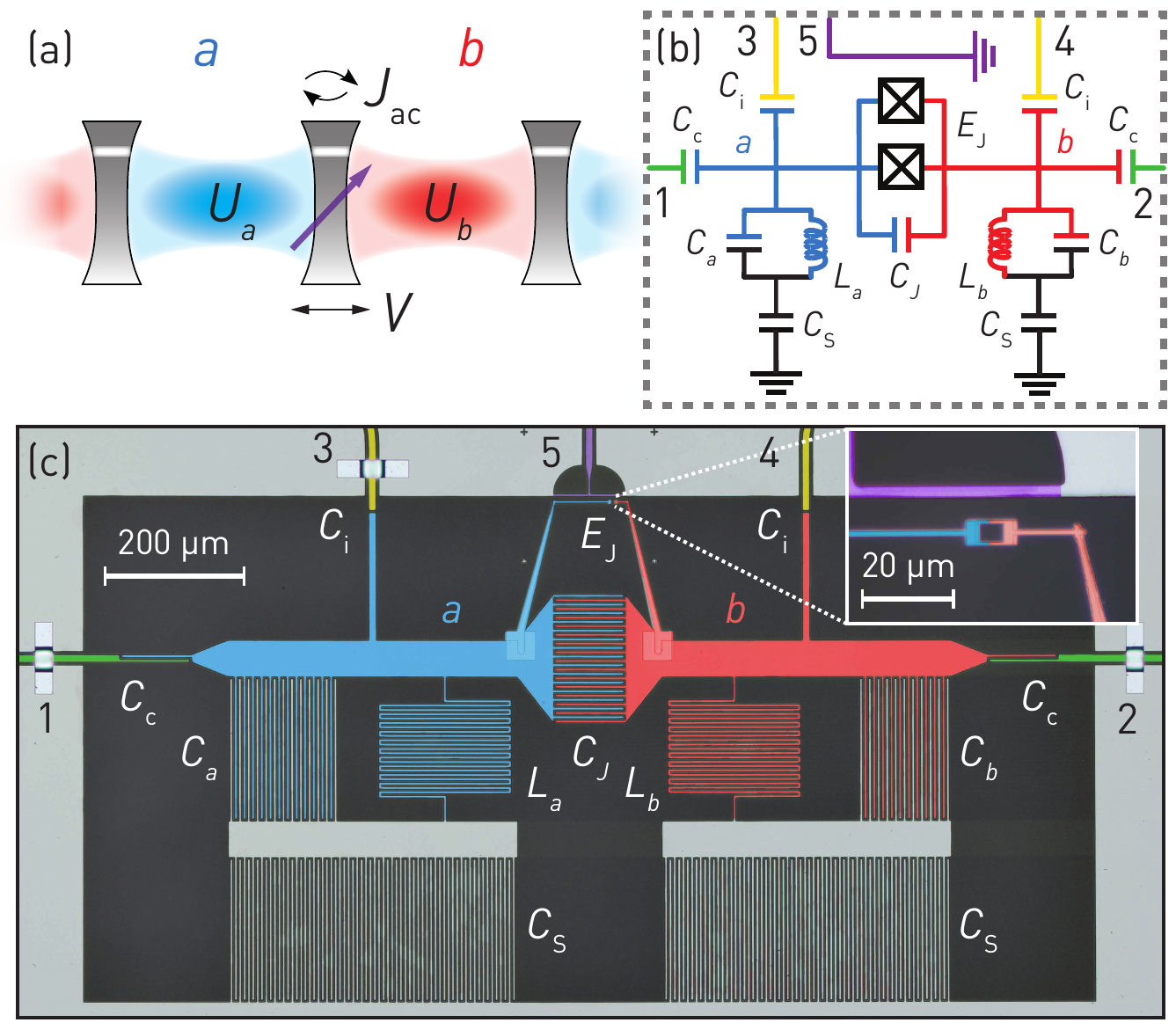}
\caption{
(a) Sketch of an optical analogue of the setup, consisting of two resonator modes $a$ and $b$ with Kerr nonlinearities $U_a$ and $U_b$, coupled via a cross-Kerr interaction $V$ and a tunable linear hopping rate $J\ind{ac}$.
(b) Equivalent circuit diagram and (c) false-colored micrograph of the sample, featuring two lumped-element $LC$ resonators $a$, $b$ (blue, red), coupled via a nonlinear coupling element composed of a capacitor and a SQUID (inset). The resonators are accessed via two symmetric sets of weakly coupled input lines (yellow, ports 3 and 4) and output lines (green, ports 1 and 2). A flux modulation tone is applied via a dedicated T-shaped flux line (purple, port 5).}
\label{fig1}
\end{figure}
%%%%%%%%%%%

Non-equilibrium coupled resonator systems have more recently also been investigated experimentally, both in a
semiclassical and in a quantum regime. Macroscopic self trapping of exciton polaritons has been observed in a dimer of coupled Bragg stack microcavities \cite{abbarchi_macroscopic_2013}, vacuum squeezing was demonstrated in a dimer of superconducting resonators \cite{eichler_quantum-limited_2014}, the unconventional photon blockade has been observed in the optical and the microwave domain \cite{vaneph_observation_2018, snijders_observation_2018}, and signatures of bistability have been found in a chain of superconducting resonators \cite{fitzpatrick_observation_2017}. Moreover, a transition from a classical to a quantum regime has been observed in the decay dynamics of a resonator dimer \cite{raftery_observation_2014}, chiral currents of one or two photons have been generated in a three qubit ring \cite{roushan_chiral_2017}, and spectral signatures of many-body localization \cite{roushan_spectroscopic_2017} as well as a Mott insulator of photons \cite{ma_dissipatively_2018} have been observed in a qubit chain.

In this Letter, we explore the interaction between individual photons in a driven-dissipative system of two nonlinearly coupled superconducting resonators (see Fig.~\ref{fig1}a). The nonlinear coupler mediates a cross-Kerr interaction $V$, on-site Kerr interactions $U_a$ and $U_b$, and an effective linear hopping interaction {with} \textsl{in-situ} tunable rate $J\ind{ac}$. We measure the on-site  $g^{(2)}_{aa} := g^{(2)}_{aa}(\tau = 0)$ and cross correlations at zero time delay $g^{(2)}_{ab} := g^{(2)}_{ab}(\tau = 0)$ between the emitted field from both resonators.
In the limit of small $J\ind{ac}$/$V$, a photon trapped in one resonator blocks the excitation of the neighboring resonator and vice versa, leading to a spontaneous self-ordering of microwave photons \cite{chang_crystallization_2008, hartmann_polariton_2010}.
Such an inter-site photon blockade regime has been predicted for resonator arrays with nonlinear couplers \cite{jin_photon_2013, jin_steady-state_2014}.
When increasing $J\ind{ac}/V$, however, a delocalization of photons and a simultaneous occupation of both resonators becomes favorable, leading to a change in the photon statistics.

For this experiment we utilize an on-chip superconducting circuit consisting of two lumped element resonators with characteristic impedance $Z  = 80 \unit{\Omega}$ (see Fig. \ref{fig1}b,c).
The aforementioned nonlinear coupling circuit, interconnecting the two resonators, consists of a capacitively shunted superconducting quantum interference device (SQUID) with capacitance $C_J = 95 \unit{fF}$ and Josephson energy $E\ind{J}\dni{max}/h = 80 \unit{GHz}$, with the Planck constant $h$.
We use a superconducting coil and an on-chip flux drive line (port 5) to ensure full dc and ac control of the magnetic flux threading the SQUID loop.
Each resonator is weakly coupled
to an input port (3 and 4), through which we drive the system, and to an output port (1 and 2)
into which approximately $50\%$ of the intra-cavity field is emitted and measured using a linear detection chain. The total decay rates are measured to be $(\kappa_a, \kappa_b)/2\pi = (2.8, 2.4) \unit{MHz}$.

\begin{figure}[t] % !b H
\centering
\includegraphics[width = 0.48\textwidth]{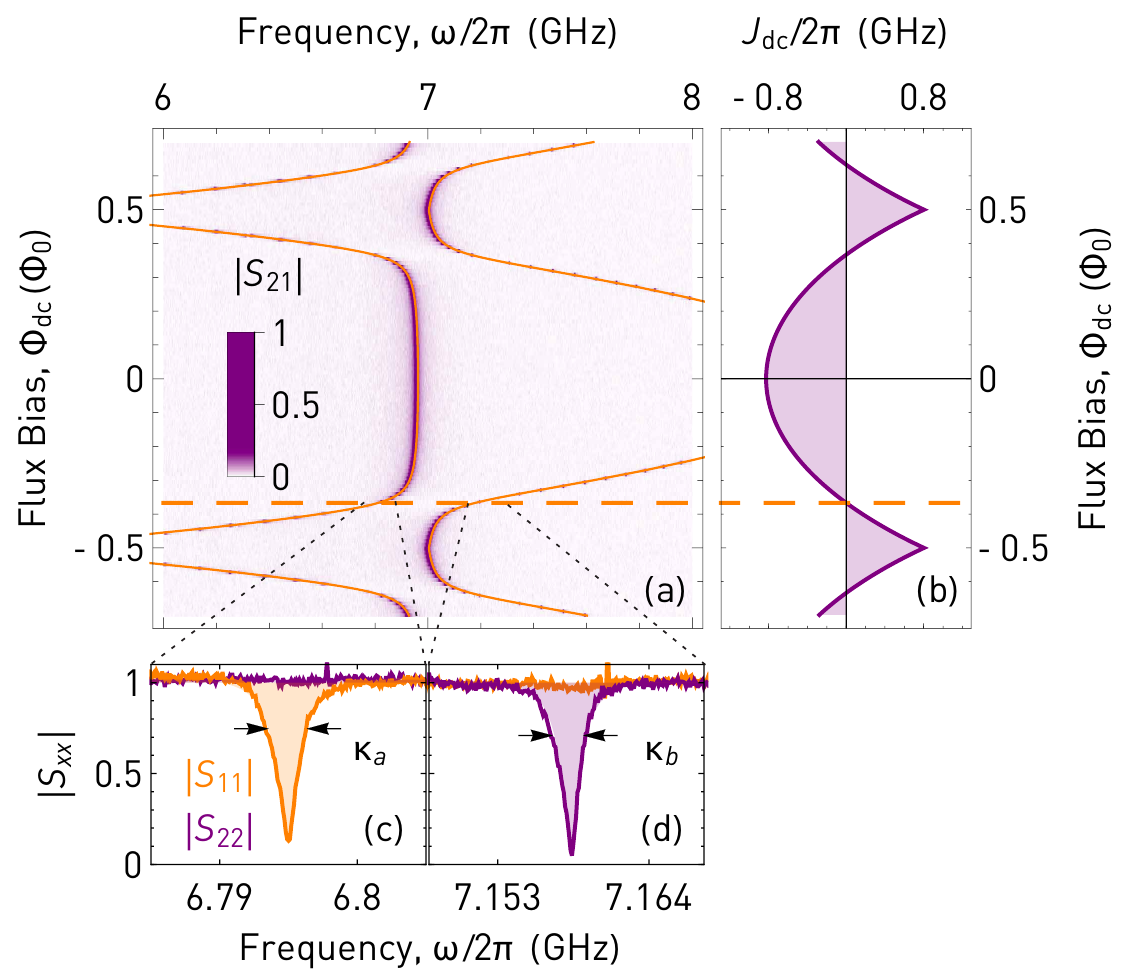}
\caption{
(a) Measured transmission amplitude $|S_{21}|$ \textsl{vs.} magnetic flux $\Phi\ind{dc}$ and fit of the resonance frequencies to a linear circuit impedance model (thin orange line). The working point $J\ind{dc} = 0$ is indicated by a dashed orange line.
(b) Linear hopping rate $J\ind{dc}$ \textsl{vs.} $\Phi\ind{dc}$, calculated using a normal mode model based on the circuit parameters extracted from (a).
(c), (d) Reflection coefficient measurements of bare cavity modes at $\Phi\ind{dc} \approx -0.37\, \Phi_0$ ($J\ind{dc} \approx 0$) with fit to a Lorentzian (solid filling).
}
\label{supp_cal1}
\end{figure}

First, we characterize the sample by measuring the transmitted amplitude $|S_{21}|$ as a function of external magnetic flux $\Phi\ind{dc}$. At each flux bias point we observe two resonances corresponding to the two eigenmodes of the system, see Fig. \ref{supp_cal1}a. The flux dependence of the measured eigenfrequencies is well explained by a linear circuit impedance model comprising a tunable effective Josephson energy, which allows us to determine the aforementioned circuit parameters.
From a normal mode model 
we extract the tuning range of the corresponding linear hopping rate $J\ind{dc}/2\pi = -0.8\ldots0.8\unit{GHz}$ (Fig. \ref{supp_cal1}b).
The tunability of $J\ind{dc}$ results from an interplay between the capacitive and the flux-dependent inductive coupling between the two resonators. As these carry opposite signs, we are able to cancel both contributions achieving approximately zero net static linear coupling $J\ind{dc} \approx 0$
at a dc flux bias point of $\Phi\ind{dc} \approx -0.37\unit{\Phi_0}$, where $\Phi_0 = \frac{h}{2 e}$ is the magnetic flux quantum. At this bias point the two measured resonances $(\omega_a, \omega_b)/2\pi = (6.802, 7.164) \unit{GHz}$ are separated by the bare detuning $\Delta/2\pi \equiv (\omega_b- \omega_a)/2\pi = 362 \unit{MHz}$ and correspond to good approximation to the local modes of the system (Fig.~\ref{supp_cal1}c,d). As a result,
the radiation of each mode ($a$, $b$) is collected in its respective output line at port (1, 2).
Notably, the finite detuning $\Delta$ between the bare cavity modes suppresses undesired nonlinear interactions, which would otherwise give rise to pair hopping and correlated hopping (see supplementary material).

\begin{figure}[t] % !b H
\centering
\includegraphics[width = 0.48\textwidth]{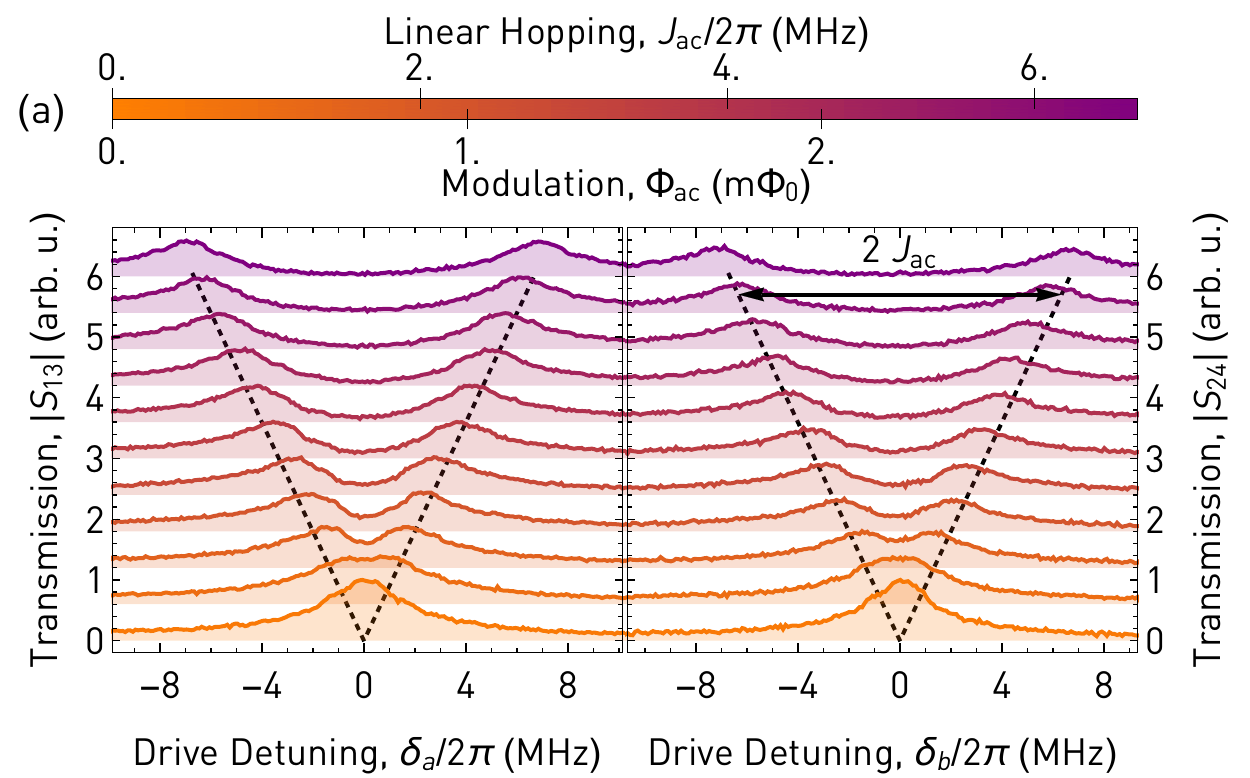}

\vspace{0.3cm}

\includegraphics[width = 0.48\textwidth]{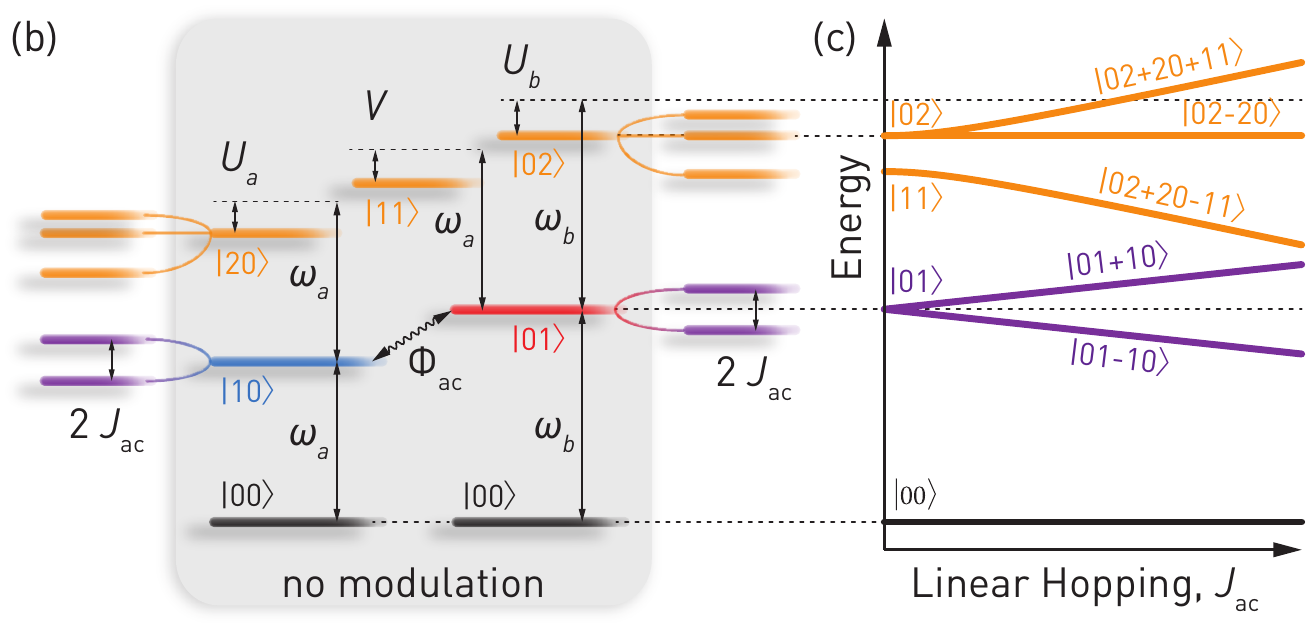}
\caption{
(a) Measured transmission amplitude $|S\ind{13}|$ ($|S\ind{24}|$) through resonator $a$ ($b$) for varying flux modulation amplitude $\Phi\ind{ac}$ applied to port 5. Linear fits to the resonance frequencies of the $J\ind{ac}$-hybridized modes are shown as black dashed lines.
(b) Energy level diagram for vanishing (gray box) and finite linear hopping rate $J\ind{ac}$ via parametric modulation at the frequency difference $\Delta = \omega_b - \omega_a$.
(c) Energy levels of resonator $b$ in the first (purple) and second (orange) excitation manifold \textsl{vs.} $J\ind{ac}$.
}
\label{fig2}
\end{figure}

In order to recover a well controllable linear hopping rate despite the finite cavity detuning, we implement a parametric coupling scheme \cite{bertet_parametric_2006, tian_parametric_2008, lu_universal_2017}. Here, we apply an ac modulated flux drive to the SQUID with a variable amplitude $\Phi\ind{ac}$ and a modulation frequency $\omega\ind{ac}$, which equals the resonator detuning $\omega\ind{ac}=\Delta$.
For $\Phi\ind{ac}=0$ we recover the uncoupled resonator modes when probing the transmission spectra $|S_{13}|$  and $|S_{24}|$ (see Fig.~\ref{fig2}a). However, as we increase $\Phi\ind{ac}$, we observe a simultaneous frequency splitting of both modes, which scales linearly with $\Phi\ind{ac}$, and which we interpret as the result of a parametrically induced photon hopping with rate $J\ind{ac}/2\pi = 0\ldots40 \unit{MHz}$ (see supplementary material).

In an appropriate doubly rotating frame, where each mode rotates at its resonance frequency, our system is well described by an effective Hamiltonian
\begingroup
%\allowdisplaybreaks
\begin{align*}
\frac{1}{\hbar}\mathcal{H}_\Delta &= \delta_a {\color{black}a\dag a} + \delta_b {\color{black}b\dag b} +  J\ind{ac}\left( {\color{black}a\dag b}+ {\color{black}b\dag a}\right)\\
&+ \frac{1}{2} U_a \, {\color{black}a\dags a^2} + \frac{1}{2} U_b \,{\color{black}b\dags b^2} +  V\, {\color{black}a\dag a b\dag b}\\
&+ \Omega_a (a\dag + a) + \Omega_b (b\dag + b)
%\\&
\end{align*}%
\endgroup

\noindent with the drive detuning $\delta_i = \omega\ind{drive,\textit{i}} - \omega_i$ ($i \in \{a,b\}$) and the drive rates $\Omega_i$. The on-site and the cross-Kerr interaction rates at zero coupling bias are $(U_a, U_b, V)/2\pi = -(3.1\pm0.3, 2.7\pm0.2, 7.0\pm0.3) \unit{MHz}$, which have been extracted from a spectroscopic measurement (see supplementary material).
In the absence of a parametric modulation the eigenstates of this Hamiltonian correspond to the photon number states $|n_a n_b\rangle$ in the local basis (compare Fig.~\ref{fig2}b). The second order transitions are red shifted by the corresponding Kerr rates. For finite $J\ind{ac}$ the eigenstates hybridize in both the one- and two-photon manifold.

We focus on a parameter regime in which $|V|$ and $J\ind{ac}$, as well as $\kappa_i$ and $\Omega_i$, are comparable in magnitude, featuring a competition between nonlinear interaction and linear hopping, as well as between drive and dissipation. In our system we additionally have $|U_i| \approx \kappa_i$. Both  $\Omega = \Omega_a = \Omega_b$, setting the average number of excitations in the system, and $J\ind{ac}$, setting the rate at which the resonators exchange excitations, are utilized as tunable control parameters, while $V$, $U_i$ and $\kappa_i$ are constant.
\@ In the experiment we keep the drive frequencies, and thus $\delta_i = 0$, fixed.
We eliminate influences of the phase of $J\ind{ac}$ on the measured results by averaging over multiple randomized phase configurations.

\begin{figure}[t] % !b H
\centering
\includegraphics[width = 0.48\textwidth]{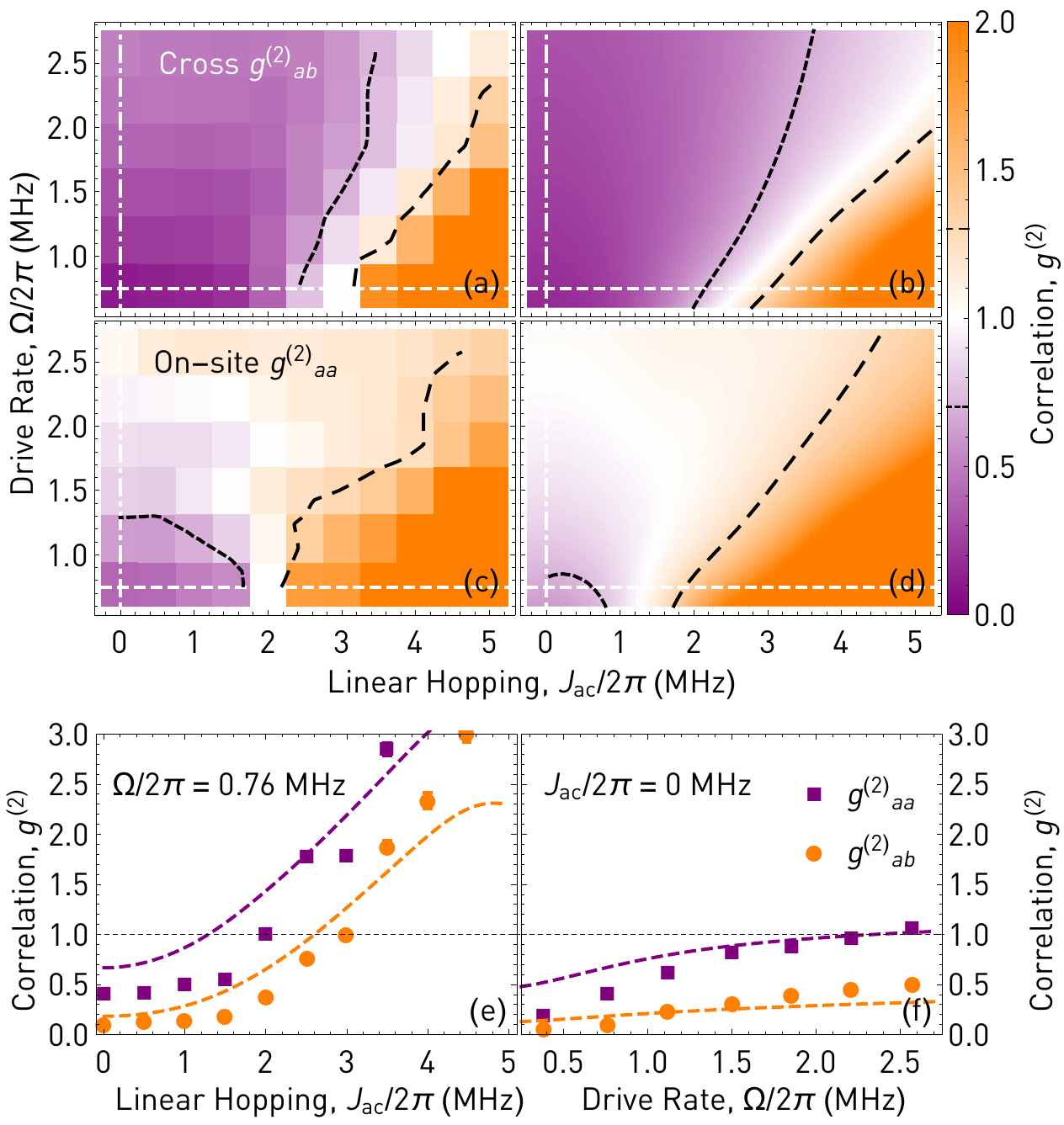}
\caption{
(a), (c) Measured cross (on-site) second order photon correlator $g^{(2)}_{ab}$ ($g^{(2)}_{aa}$) as a function of the linear hopping rate $J\ind{ac}$ and the drive rate $\Omega$; the black dotted (dashed) line indicates the linearly interpolated contour for $g^{(2)} = 0.7$ ($g^{(2)} = 1.3$).
(b), (d) Corresponding results from numerical simulations.
(e) $g^{(2)}$ \textsl{vs.} $J\ind{ac}$, cut for $\Omega/2\pi = 0.76 \unit{MHz}$ (see white dashed line; measured data is shown with markers, numerical simulations with dashed lines).
(f) $g^{(2)}$ \textsl{vs.} $\Omega$, cut for $J\ind{ac}/2\pi = 0 \unit{MHz}$ (see white dash-dotted line).
}
\label{fig3}
\end{figure}

We characterize the quantum states of the uniformly and continuously driven two-resonator system by measuring the second order cross $g^{(2)}_{ab}$ and on-site correlation $g^{(2)}_{aa}$ of the emitted radiation as a function of $J\ind{ac}$ and  $\Omega$ (see Fig. \ref{fig3}a,c). To this aim, we linearly amplify and digitize the radiation fields at both output ports in order to obtain the second order photon correlations \cite{bozyigit_antibunching_2011, eichler_quantum-limited_2014, eichler_exploring_2015}. To enhance the signal-to-noise ratio, we use a quantum-limited Josephson parametric amplifier \cite{eichler_quantum-limited_2014} operated in a phase-sensitive mode (see supplementary material for details about the detection process).
The measured $g^{(2)}$ correlations are compared with the results of a numerical master equation simulation \cite{qutip} (see Fig.~\ref{fig3}b,d).
As confirmed by this simulation, the average resonator occupations remain at or below the single photon level for all the data presented in Fig.~\ref{fig3}.

In the regime of small $J\ind{ac}$ and low $\Omega$ we measure the radiation to be anti-bunched, see Fig.~\ref{fig3}.
In this limit, the cross-Kerr interaction effectively shifts the transition frequency of one cavity when a photon is present in the other and thus detunes the ($|01\rangle, |10\rangle \leftrightsquigarrow |11\rangle$) transition from the drive tones. This inhibits simultaneous occupation of both cavities, leading to a dynamic self-ordered photon state manifested as anti-bunching in the photon cross statistics.
Equivalently, the on-site Kerr interaction prevents each mode from being doubly excited, leading to anti-bunched on-site correlations.

Increasing the hopping rate $J\ind{ac}$
results
in a hybridization of the modes in both the one- and two-excitation manifold,
see Fig.~\ref{fig2}c for a level diagram as a function of $J\ind{ac}$.
When $J\ind{ac}$ becomes comparable to the Kerr rate $U_i$, the transitions of the single photon manifold become detuned from the drive frequency, while the two-photon-transition into the symmetric $|02+20+11\rangle$ branch becomes resonant with the drive.
This leads to a more efficient drive into the second excitation manifold compared to the originally dominating single photon states and to an admixture of simultaneous cavity occupations ($|11\rangle$, $|20\rangle$, and $|20\rangle$). This causes a crossover from anti-bunched to bunched statistics in both the measured $g^{(2)}_{ab}$ and $g^{(2)}_{aa}$, see Fig.~\ref{fig3}e.
Interestingly, we find a regime in which the on-site correlation $g^{(2)}_{aa}$ is already close to unity, while the cross correlation $g^{(2)}_{ab}$ is still anti-bunched. We attribute this effect to $V$ being larger than $U_i$.

Studying the dependence on the drive rate $\Omega$, we find that $g^{(2)}_{aa}$ approaches unity when $\Omega$ exceeds $U_i$ (Fig. \ref{fig3}f), which we explain by the breakdown of the photon blockade. This effect is found to be largely independent of $J\ind{ac}$. We observe a similar behavior for the cross correlations. In this case, however, the measured $g^{(2)}_{ab}$ approaches one half in the limit of large drive rate $\Omega$, which is in good agreement with the result obtained from the numerical simulations.

In conclusion, we have realized a coupled cavity system, featuring a tunable ratio between linear hopping and cross-Kerr interaction rate and observed the crossover from photon ordering to delocalization. Inspired by the proposals by Jin \textsl{et al.}  \cite{jin_photon_2013, jin_steady-state_2014}, we interpret the measured cross correlations as an order parameter in a ($J\ind{ac}$, $\Omega$)-dependent phase diagram of the system. The observed crossover closely resembles the onset of a driven-dissipative photon ordering phase transition, from a fully ordered crystalline phase dominated by spontaneous symmetry breaking towards a uniform delocalized steady-state phase \cite{brown_localization_2018, fink_signatures_2018}.

We expect the demonstrated coupling mechanism to be well extendable towards larger resonator arrays. Resilience to disorder in electrical parameters \cite{underwood_low-disorder_2012} can be achieved by frequency staggering of neighboring cavities along with the adjustability of the parametric modulation frequencies. Additionally, the employed lumped element structures excel in this scenario thanks to a compact footprint and high design versatility.

The presented system and variations thereof could be used to explore regimes, in which inter-site interactions exceed on-site interactions \cite{elliott_designing_2018, busche_contactless_2017}. Additionally, the controllability of the phase of the hopping rate could be employed to create artificial gauge fields in plaquette systems and to study non-reciprocal dynamics with photons \cite{roushan_chiral_2017}. Furthermore, the variability of flux modulation frequencies could enable the controllable activation of additional interaction terms such as a parametric coupling between neighboring resonators \cite{tangpanitanon_hidden_2018} or pair hopping \cite{peropadre_tunable_2013}, \textsl{e.g.} for the study of
supersolid phases \cite{huang_extended_2016}.

This  work  is  supported  by the National Centre of Competence in Research ``Quantum Science and Technology'' (NCCR QSIT), a research instrument of the Swiss National Science Foundation (SNSF) and by ETH Zurich. M.~J.~H. acknowledges support by the EPSRC under grant No. EP/N009428/1.
\putbib
%\bibliography{fftcd_library_v5}

\end{bibunit}
%%% Supplementary  %%%%%%%%%%%%%%%%%%%%%%%%%%%%%%%%%%%%%%%%
\begin{bibunit}[apsrev4-1]

\widetext
%\pagebreak
\clearpage

\setcounter{equation}{0}
\setcounter{figure}{0}
\setcounter{table}{0}
\setcounter{page}{1}
\makeatletter
\renewcommand{\theequation}{S\arabic{equation}}
\renewcommand{\thefigure}{\text{S}\arabic{figure}}
\renewcommand{\thetable}{\text{S}\Roman{table}}
\renewcommand{\bibnumfmt}[1]{[S#1]}
\renewcommand{\citenumfont}[1]{S#1}
\makeatother

\onecolumngrid

\begin{center}
\textsl{\textbf{\large{Supplementary Information to}}}
\vspace{0.05cm}

\centering\large{\textbf{\puttitle}}
\vspace{0.4cm}

\normalsize{
{Michele C. Collodo},$^{1,*}$
{Anton Poto\v{c}nik},$^1$
{Simone Gasparinetti},$^1$
{Jean-Claude Besse},$^1$
{Marek Pechal},$^{1,\dagger}$\\
{Mahdi Sameti},$^2$
{Michael J. Hartmann},$^2$
{Andreas Wallraff},$^1$ and
{Christopher Eichler}$^{1,\ddagger}$
}
\vspace{0.2cm}

\textit{\normalsize{
$^1${Department of Physics, ETH Zurich, CH-8093 Zurich, Switzerland}\\
$^2${Institute of Photonics and Quantum Sciences, Heriot-Watt University Edinburgh EH14 4AS, United Kingdom}
}}
\vspace{1cm}
\end{center}

\twocolumngrid

%\input{fftcd_supplementary_v8}
%%%%%%%%%%%%%%%%%%%%%%%%%%%%%%%%%%%%%%%%%%%%%%%%
\setlength{\ULdepth}{3.5pt}

\section{System Engineering}

\subsection{Linear Impedance Model}

We construct a linear impedance model based on the equivalent circuit of our sample (see Fig. \ref{fig1}b) using an ABCD matrix formalism. The model is then fitted to the measured eigenfrequencies, which we extract from measurements of the dc flux dependent transmission amplitude spectrum $|S_{21}|$ (see Fig. \ref{supp_cal1}a). This allows us to determine the electrical parameters, see Tab. \ref{tab:1}.

\begin{table}[h]
%\begin{ruledtabular}
\begin{tabular}{l r}
\hline
\hline
$C_a$ &	$260 \unit{fF}$\\
$C_b$ &	$300 \unit{fF}$\\
$C_J$ &	$95 \unit{fF}$\\
$E\ind{J}\dni{max}/h$$\qquad$ &$80 \unit{GHz}$	\\
$L_a$ &	$1.9 \unit{nH}$\\
$L_b$ &	$1.9 \unit{nH}$\\
$L\ind{s}$ &$0.3 \unit{nH}$\\
\hline
\hline
\end{tabular}
%\end{ruledtabular}
\caption{\label{tab:1} List of electrical parameters of the microwave circuit presented in the main text.}
\end{table}

We neglect the corrections due to the coupling to the environment. However, we take into account the contribution of a spurious inductance $L\ind{s}$ caused by the lead wires to the SQUID. This will modify the dc contribution of the effective inductance of the SQUID as $L\ind{J} = L\ind{s} + \Phi_0/E\ind{J}(\Phi\ind{dc})$ with $E\ind{J}(\Phi\ind{dc}) = E\ind{J}\dni{max} |\cos(\pi \Phi\ind{dc} / \Phi_0)|$.

In order to prevent the creation of a large closed ground loop through the SQUID, which could alter the circuit's dynamics in the presence of magnetic flux, we opt for a floating resonator configuration via large shunt capacitors, designed to be $C\ind{s} = 800 \unit{fF}$.

\subsection{Effective Hamiltonian under Parametric Modulation}

In order to be able to study the competition of linear hopping interaction and nonlinear cross-Kerr interaction between adjacent resonators, it is crucial to construct a system with a Hamiltonian featuring exclusively these two coupling mechanisms.

Starting from the Lagrangian of the implemented nonlinear coupling circuit \cite{jin_photon_2013} discussed in the main text (see Fig. \ref{fig1}), we find the full local mode Hamiltonian
\begin{align*}
\frac{1}{\hbar} \mathcal{H\dni{full}} &=
 \omega_a \, \uline{{\color{\colordc} a^{\dagger}a}} + \omega_b\, \uline{{\color{\colordc}b^{\dagger} b}} \\
&- (J\ind{c} - J_\ell) \left({\color{\colortwo}a\dags} + {\color{\colortwo}b\dags} + {\color{\colortwo}a^2} + {\color{\colortwo}b^2} \right)\\
&+ (J\ind{c} - J_\ell) \left( \dotuline{{\color{\colord}a\dag b}} + \dotuline{{\color{\colord}b\dag a}} \right)\\
&- (J\ind{c} + J_\ell) \left( {\color{\colorsum}a\dag b\dag} + {\color{\colorsum}b a} \right)\\
&+\frac{\tilde V}{24} \underbrace{\left( a + a\dag - b - b\dag\right)^4}_{
\large{
\begin{subarray}{l}
= a\dagt+ 4\,{\color{\colortwo}a\dagc a}+ 6\, \uline{{\color{\colordc}a\dags a^2}}  + 4\, {\color{\colortwo}a\dag a^3}   + a^4\\[7pt]
 - 4\, b\dag a\dagc - 4\,{\color{\colorthree}a\dagc b} - 12 \,{\color{\colorsum}b\dag a\dags a}- 12\,  \dotuline{{\color{\colord}a\dags b a}}\\
- 12\, \dotuline{{\color{\colord} b\dag a\dag a^2}}- 12\,{\color{\colorsum} a\dag b a^2}  - 4\, {\color{\colorthree}b\dag a^3}  - 4\, b a^3 \\[7pt]
+ 6\, b\dags a\dags   + 12\, {\color{\colortwo}b\dag a\dags b} + 6\, \dashuline{{\color{\colordd}a\dags b^2}}\\
+ 12\, {\color{\colortwo}b\dags a\dag a} + 24\, \uline{{\color{\colordc}b\dag a\dag b a}}+ 12\,{\color{\colortwo} a\dag b^2 a}\\
+ 6\,\dashuline{{\color{\colordd} b\dags a^2}}    + 12\, {\color{\colortwo}b\dag b a^2}   + 6\, b^2 a^2 \\[7pt]
- 4\,b\dagc a\dag - 4\,{\color{\colorthree}b\dagc a} - 12\, {\color{\colorsum}b\dags a\dag b}  - 12\,  \dotuline{{\color{\colord}b\dags b a}}\\
- 12\,  \dotuline{{\color{\colord}b\dag a\dag b^2}}   - 12\, {\color{\colorsum}b\dag b^2 a} - 4\, {\color{\colorthree}a\dag b^3 }  - 4\, b^3 a   \\[7pt]
+b\dagt + 4\,{\color{\colortwo}b\dagc b}  +6\, \uline{{\color{\colordc} b\dags b^2}}  + 4\, {\color{\colortwo}b\dag b^3} + b^4%\\[7pt]
\end{subarray}
}
}  + \mathcal{O}(a^6)\\
\end{align*}

\noindent expressed in terms of ladder operators $a^{\dagger}$, $b^{\dagger}$ with the bare resonator frequencies $\omega_a$, $\omega_b$, the capacitively (inductively) mediated linear hopping rate $J\ind{c}$ ($J_\ell$) and the cross-Kerr rate $\tilde V$.
For simplicity we assume comparable characteristic impedances $Z_a \approx Z_b$ resulting in Kerr rates $\tilde V \approx 2 \tilde U_a \approx 2 \tilde U_b$ for the two resonators. Corrections from normal ordering are omitted. The selected highlighted terms rotate at a frequency of
$
\{
\dashuline{{\color{\colordd} 2\omega_a-2\omega_b}},
\dotuline{{\color{\colord}\omega_a - \omega_b}},
\uline{{\color{\colordc}0}} \}
$
with respect to a doubly rotating frame, which is locked to the resonance frequencies of both modes $\omega_a$ and $\omega_b$.
In particular, this shows the importance of a substantial detuning $\Delta = \omega_b - \omega_a \neq 0$ in order to suppress the detrimental
pair hopping ($\sim\dashuline{{\color{\colordd}a^{\dagger 2} b^{2}}}$) and
correlated hopping ($\sim\dotuline{{\color{\colord}a^{\dagger} \left(a^{\dagger}a + b^{\dagger}b\right) b}}$) terms while keeping the desired Kerr terms ($\sim\uline{{\color{\colordc}a^{\dagger}a b^{\dagger}b}}$) resonant.

%\onecolumngrid

\begin{figure*}[t] % !b H
\centering
\includegraphics[width = 0.9\textwidth, angle=0]{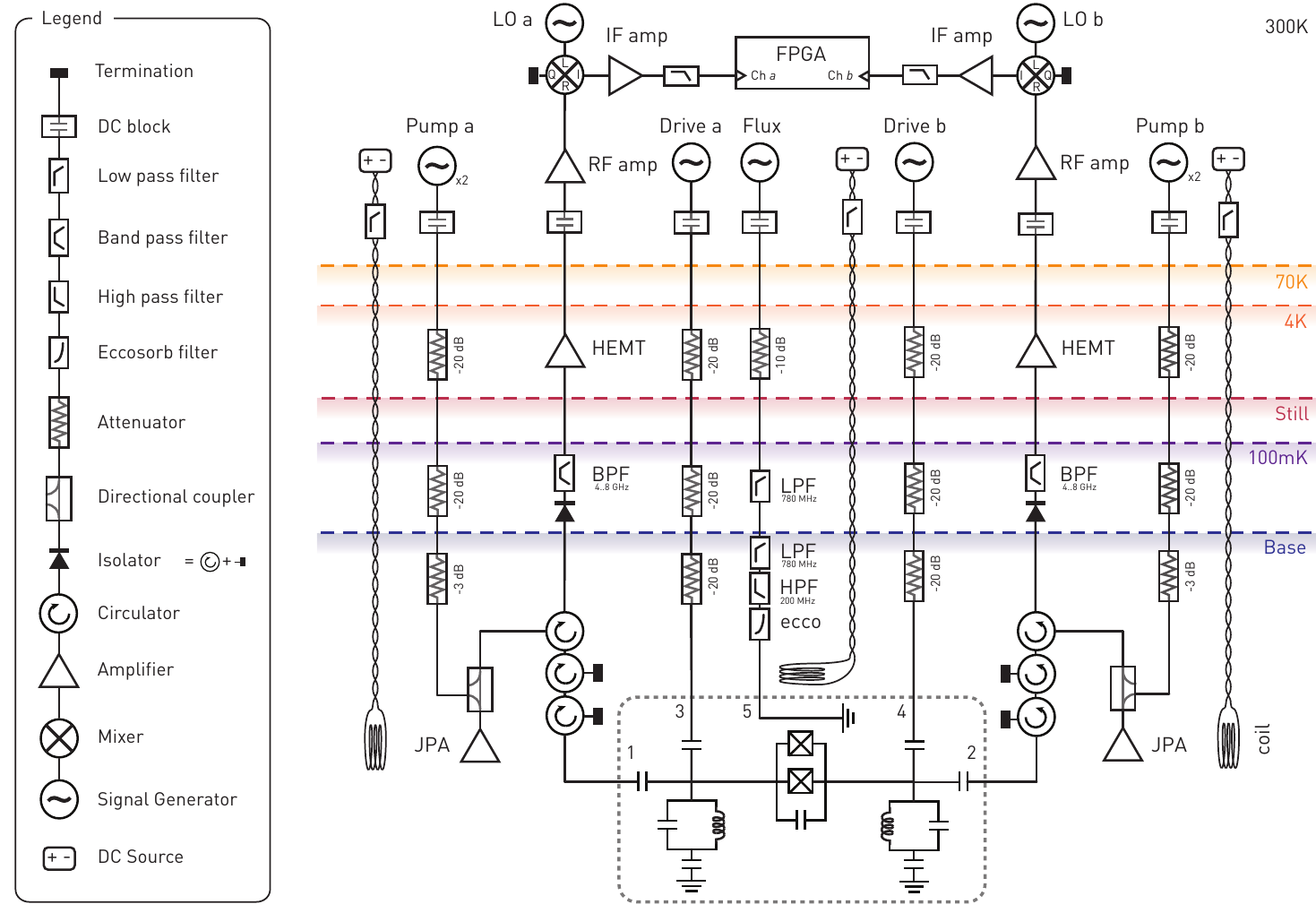}
\caption{Full wiring diagram of the experimental setup. The setup consists of two symmetrical linear amplification chains composed of reflective Josephson parametric amplifiers, HEMT amplifiers and room temperature amplifiers.}
\label{supp2}
\end{figure*}
%\twocolumngrid

In order to preserve a linear hopping interaction (${\sim\dotuline{{\color{\colord}a^{\dagger} b}}}$) despite this detuning $\Delta$ we employ a parametric modulation scheme by driving the nonlinear coupling circuit with the ac modulated flux
$$\varphi(t)~=~\varphi\ind{dc} +  \varphi\ind{ac} \cos\left(\omega\ind{ac}t \right),$$
where $\varphi_{i} = 2\pi \Phi_i / \Phi_0$ are the magnetic flux amplitudes normalized to the flux quantum. The applied flux drive tone at frequency $\omega\ind{ac}$ is responsible for selecting and activating specific interactions from $\mathcal{H\dni{full}}$ within a subsequent rotating wave approximation. The interaction rates are effectively altered as
$ J_\ell\mapsto J_\ell  \cos \frac{\varphi(t)}{2}  \text{ and } \tilde V \mapsto \tilde V \cos \frac{\varphi(t)}{2} $, an expansion to first order leads to
\begin{align*}
\cos \left(\frac{\varphi(t)}{2}\right) &= \cos \left(\frac{\varphi\ind{dc}}{2} +  \frac{\varphi\ind{ac}}{4} \left(\e{\i \Delta t} + \e{-\i \Delta t}\right) \right)\\
&\approx \cos \left(\frac{\varphi\ind{dc}}{2}\right) - \sin \left(\frac{\varphi\ind{dc}}{2}\right) \frac{\varphi\ind{ac}}{4} \left(\e{\i \Delta t} + \e{-\i \Delta t}\right)
\end{align*}
for a modulation frequency $\omega\ind{ac} = \Delta$.

We choose individual rotating frames for each mode, locked to their respective resonance frequency $\omega_{a}$, $\omega_{b}$.
After a rotating wave approximation for sufficiently large resonator-resonator detuning $\Delta$ (\textsl{i.e.} keeping solely resonant terms) and operating with balanced capacitive and inductive hopping rates $J\ind{c} = J_\ell \cos (\varphi\ind{dc}/2)$, we are left with the effective Hamiltonian

\begin{align*}
\frac{1}{\hbar}\mathcal{H\dnis{i}\inds{$\Delta$}} &= \delta_a \uline{{\color{\colordc}a\dag a}} + \delta_b \uline{{\color{\colordc}b\dag b}}\\
&+ J_\ell \sin \left(\frac{\varphi\ind{dc}}{2}\right)  \frac{\varphi\ind{ac}}{4} \left( \dotuline{{\color{\colord}a\dag b}}+ \dotuline{{\color{\colord}b\dag a}}\right)\\
&+\frac{1}{2} \tilde  U_a \cos  \left(\frac{\varphi\ind{dc}}{2}\right)\, \uline{{\color{\colordc}a\dags a^2}}\\
&+ \frac{1}{2} \tilde  U_b   \cos  \left(\frac{\varphi\ind{dc}}{2}\right)\,\uline{{\color{\colordc}b\dags b^2}} \\
&+ \tilde V \cos  \left(\frac{\varphi\ind{dc}}{2}\right) \, \uline{{\color{\colordc}a\dag a b\dag b}}\\
&- \tilde J_n \sin  \left(\frac{\varphi\ind{dc}}{2}\right)  \frac{\varphi\ind{ac}}{4} \left( \dotuline{{\color{\colord}a\dag \left(a\dag a + b\dag b \right) b}} + \text{h.c.}\right)\\
&+ \Omega_a (a\dag + a) + \Omega_b (b\dag + b)
\end{align*}

In this form, the linear scaling of the effective linear hopping rate $J\ind{ac} = J_\ell \sin(\varphi\ind{dc}/2) \, \varphi\ind{ac}/4$ with the flux modulation amplitude $\Phi\ind{ac}$ is evident. For our device we find $J_\ell/2\pi = 1.6 \unit{GHz}$.
The nonlinear Kerr rates
$(U_a, U_b, V) = (\tilde U_a, \tilde U_b, \tilde V) \cos (\varphi\ind{dc}/2)$
are dependent on the flux bias point, but are inherently robust against detuning $\Delta$ or flux modulation.
We infer the correlated hopping rate $\tilde J_n/2\pi = 8 \unit{MHz} $ from the measured cross-Kerr rate. For sufficiently weak flux modulation ($\varphi\ind{ac}/2\pi \lesssim 3\times10^{-3}$ in the experiment), the contribution of the pair (correlated) hopping terms are quadratically (linearly) suppressed, allowing us to neglect their influence on the system dynamics and to finally reconstruct the Hamiltonian as presented in the main text.

\section{Experimental Setup and Data Collection}

\subsection{Sample Fabrication}

All linear elements of the device presented in Fig. \ref{fig1} are fabricated by patterning a sputtered $150 \unit{nm}$ thin niobium film on a sapphire substrate with photolithography and reactive ion etching.
In a subsequent step Josephson junctions are added using electron-beam lithography and double-angle shadow evaporation of Aluminum.
We operate the sample in a dilution refrigerator at a base temperature of $20 \unit{mK}$.
The full microwave wiring diagram is shown in Fig. $\ref{supp2}$.

\subsection{Linear Amplification Chain and Driving Scheme}

The resonators $a$, $b$ are driven at their respective bare resonance frequencies $\omega_a$, $\omega_b$ with a set of symmetric drive lines via ports 3 and 4. Scattered radiation is collected at ports 1 and 2 and routed to a set of symmetric linear amplification chains. The itinerant signal is amplified by near-quantum limited Josephson parametric amplifiers (JPA) \cite{eichler_quantum-limited_2014} at base temperature,
operated in phase-sensitive mode in a dual pump tone configuration \cite{kamal_signal--pump_2009}.
The pump tones are symmetrically detuned from the signal frequency by $\pm 250 \unit{MHz}$ ($\pm 300 \unit{MHz}$) for the amplifier at mode $a$ ($b$). The pump tones are thus far detuned from the measurement band, alleviating the need for pump tone cancellation.
We determine a bandwidth of $13 \unit{MHz}$ ($20 \unit{MHz}$) at a gain of $18.8 \unit{dB}$ ($19.6 \unit{dB}$) of the parametric amplifiers.

Subsequently, the signals are amplified by high-electron-mobility-transistor (HEMT) amplifiers, thermally anchored at $4\unit{K}$, and low noise room temperature amplifiers. The signals are then down converted by mixing with individual local oscillator tones to an intermediate frequency of $25 \unit{MHz}$, filtered to avoid backfolding of noise and amplified before being digitized by an analog-to-digital converter with a sampling rate of $100 \unit{MHz}$.
A field programmable gate array (Xilinx Virtex-4) digitally down converts the digitized signals and extracts the $I$, $Q$ quadratures for each channel.

The JPA pump field and local oscillator phases are locked to each other and chosen such that $I$ corresponds to the amplified quadrature. In order to uniformly sample the entire phase space distribution of the field we slowly cycle the relative phase of the individual local oscillators with respect to the corresponding input drive field. Consequently, we are able to reconstruct the second order correlations from a measurement of a single quadrature $I$ per channel.

\subsection{Correlation Function Measurements}

We collect the resulting $I_a$, $I_b$ quadrature values of several million repetitions of the experiment with the drive fields turned on in a two dimensional histogram (``on'') and extract the normally ordered statistical moments $\langle (I_a)^n (I_b)^m\rangle\ind{on}$ (with $n+m \leq 4$) of this distribution \cite{lang_correlations_2013}.
It is necessary to mitigate the influence of added thermal noise on the on-histogram in order to reconstruct the statistics of the radiation field emitted from the sample at base temperature.
To this aim, we repeat the measurement without applying any drive field and extract the statistical moments $\langle (I_a)^n (I_b)^m\rangle\ind{off}$ from the corresponding histogram (``off'').
Assuming a linear amplification chain, the moments of the on-histogram
\begin{align*}
\langle (I_a)^n (I_b)^m\rangle\ind{on} = \sum^{n,m}_{k,l = 0} \binom{n}{k}\binom{m}{l} &\langle (I_a)^{n-k} (I_b)^{m-l}\rangle\ind{off} \\
\times&\langle (I_a)^{k} (I_b)^{l}\rangle\ind{diff}
\end{align*}
are composed of the added noise captured in the moments of the off-histogram and the moments of the signal at the output of the sample $\langle (I_a)^{k} (I_b)^{l}\rangle\ind{diff}$. This set of linear equations is solved in order to obtain the latter \cite{eichler_characterizing_2012}.

The reconstructed moments of the quadratures can be expressed as normally ordered moments of the mode operators $\langle (I_a)^k (I_a)^l\rangle\ind{diff} \propto \langle:\left(a + a\dag\right)^k \left(b + b\dag\right)^l :\rangle$, where $I_a =  \frac{1}{2} \left(a\dag + a\right)$, $I_{b} = \frac{1}{2}\left(b\dag + b\right)$.
This allows us to calculate the correlations at zero time delay
$g^{(2)}_{aa} := g^{(2)}_{aa}(0) = \frac{\langle a\dag a\dag a a \rangle}{\langle a\dag a\rangle^2}$ and $g^{(2)}_{ab} := g^{(2)}_{ab}(0) = \frac{\langle a\dag a b\dag b \rangle}{\langle a\dag a\rangle \langle b\dag b\rangle}$ as
$$g^{(2)}_{aa} = \frac{2}{3} \frac{\langle (I_a)^4 \rangle\ind{diff}}{\langle (I_a)^2\rangle^2\ind{diff}}
\text{, }
g^{(2)}_{ab} = \frac{\langle (I_a)^2 (I_b)^2 \rangle\ind{diff}}{\langle (I_a)^2\rangle\ind{diff} \langle (I_b)^2\rangle\ind{diff}}$$
Error bars shown in Fig.~\ref{fig3} of the main text are extracted from the standard deviation of the mean of repeated measurements.
The validity of the analysis is verified by measuring the statistics of a coherent tone $g^{(2)}_{aa} = 1.01 \pm 0.01$ and $g^{(2)}_{bb} = 1.0 \pm 0.02$.

\section{Measured Kerr Rates}

\begin{figure}[t] % !b H
\centering
\includegraphics[width = 0.48\textwidth]{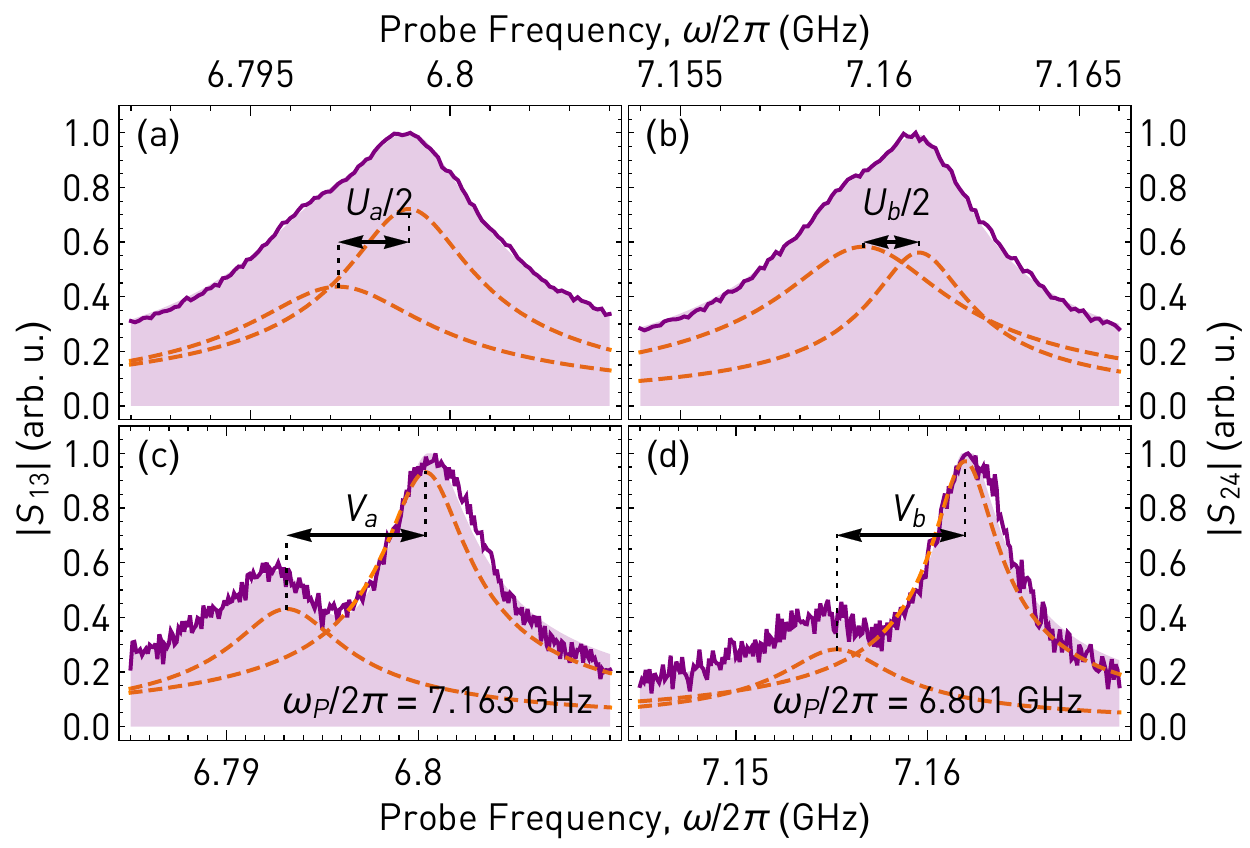}
\caption{(a) Transmitted amplitude $|S_{13}|$ (purple line), including fit to double-Lorentzian (purple filling, individually shown via orange dashed line).
(b) Transmitted amplitude $|S_{24}|$.
(c) Transmitted amplitude $|S_{13}|$ measured in two-tone spectroscopy, pump tone with frequency $\omega\ind{P}$ applied at port 4.
(d) Transmitted amplitude $|S_{24}|$ with pump tone at port 3.}
\label{supp_kerr}
\end{figure}

We measure the nonlinear interaction rates at the dc flux bias point $\Phi\ind{dc} \approx -0.37\unit{\Phi_0}$, \textsl{i.e.} at vanishing linear hopping $J\ind{dc}/2\pi \approx 0 \unit{MHz}$  via spectroscopic characterization of two-photon transitions (Fig. \ref{supp_kerr}).
The on-site Kerr rates are measured with a strong probe tone, we find $(U_a, U_b)/2\pi = -(3.1\pm0.3, 2.7\pm0.2) \unit{MHz}$.
The cross-Kerr rates are measured using a weak probe tone while simultaneously pumping the respective single photon transitions strongly. Fitting two Lorentzian curves to the measured spectra allows us to extract the detuning between the transitions of the one- and two-photon manifolds, which directly corresponds to the Kerr rates in question. We find the cross-Kerr rate $(V\ind{a}, V\ind{b})/2\pi = - (7.2, 6.7)\unit{MHz}$ to be dependent on the port from which the transition is pumped. We attribute this to imprecisions in the frequency extraction and combine the findings to the value reported in the main text $V/2\pi = - (7.0\pm0.3) \unit{MHz}$.

%%%%%%%%%%%%%%%%%%%%%%%%%%%%%%%%%%%%%%%%%%%%%%%%

%\bibliographystyle{supp}{abbrv}
%\bibliography{fftcd_library_v5}
\putbib

\end{bibunit}
\end{document}